# Permutationless many-jet event reconstruction with symmetry preserving attention networks

Michael James Fenton,[1,*] Alexander Shmakov,[2] Ta-Wei Ho,[3] Shih-Chieh Hsu,[4] Daniel Whiteson,[1] and Pierre Baldi[2]

[1]*Department of Physics and Astronomy, University of California Irvine, Irvine, California 92697, USA*
[2]*Department of Computer Science, University of California Irvine, Irvine, California 92697, USA*
[3]*Department of Physics, National Tsing Hua University, Hsinchu City 30013, Taiwan*
[4]*Department of Physics and Astronomy, University of Washington, Seattle, Washington 98195-4550, USA*

Top quarks, produced in large numbers at the Large Hadron Collider, have a complex detector signature and require special reconstruction techniques. The most common decay mode, the "all-jet" channel, results in a 6-jet final-state which is particularly difficult to reconstruct in $pp$ collisions due to the large number of permutations possible. We present a novel approach to this class of problem, based on neural networks using a generalized attention mechanism, that we call symmetry preserving attention networks (SPA-NET). We train one such network to identify the decay products of each top quark unambiguously and without combinatorial explosion as an example of the power of this technique. This approach significantly outperforms existing state-of-the-art methods, correctly assigning all jets in 80.7% of 6-jet, 66.8% of 7-jet, and 52.3% of ≥8-jet events respectively.

## I. INTRODUCTION

At the Large Hadron Collider (LHC), protons are collided at the highest energy ever produced in the laboratory. Many of these collisions produce a high multiplicity of *jets*; collimated sprays of particles that originate from the strongly coupled quarks and gluons inside the proton. Final-states containing only jets occur through a number of physical processes, and the copious production of these "all-jet" topologies presents opportunities for precision measurements [1,2], and searches for rare Standard Model [3] or new physics [4,5] processes, but also raises particular challenges. Specifically, it is typically difficult to connect an observed jet with its quark origin, and the factorial dependence on the number of jets leads to the so-called "combinatorial explosion." For example, top quark pair production with subsequent hadronic $W$-boson decays $t \to Wb \to qqb$ has a $\geq 6$ jet final-state in the so-called "resolved" regime in which each of the three decay products produced from each top are reconstructed as a single jet. In some events this can be mitigated using "boosted" reconstruction [6], though this is limited to a small subset of events [7]. Thus, one of the biggest obstacles to extracting physical information from these events is correctly determining which jets originate from each of the parent top quarks.

The top quark, as the most massive fundamental particle in the Standard Model, is the only quark to decay before hadronization. This presents a unique opportunity to study an isolated quark—if its decay products can be correctly identified. Top quarks decay almost exclusively via $t \to Wb$ in the Standard Model, and events are categorized by the decay modes of the $W$ bosons: dileptonic (9%), single-lepton (45%), or all-jets (46%) [8]. To date, the most precise measurements of top quark properties are typically performed in the single-lepton or dilepton channels [9]. The all-jet channel, held back by the ambiguous event reconstruction and large backgrounds, is comparatively underexplored.

In this paper, we propose a novel architecture for assignment of particle origin to jets, symmetry preserving attention networks (SPA-NET). Applying attention networks that naturally reflect the permutation symmetry of the task, SPA-NET significantly outperforms existing state-of-the-art techniques while avoiding combinatorial explosion. In the following, we define the nature of the jet assignment task, describe invariance and attention mechanisms in neural networks, describe the dataset and training, and demonstrate the performance of our technique relative to the state-of-the-art.

## II. JET ASSIGNMENT

The jet assignment task is the identification of the original particle which leads to a reconstructed jet. In a collision which produces $N$ jets, there are $N!$ possible assignments. Fortunately, symmetries can reduce this number.

---

*mjfenton@uci.edu

Top quarks decay via the chain $t \to Wb \to qqb$, suppressing charge labels. A pair of top quarks, $tt'$, therefore produce six quarks, $qqbq'q'b'$. This process is shown in Fig. 1. The task is to correctly identify six observed jets with six labels: $b$, $b'$, $2 \times q$, and $2 \times q'$. The symmetries between the two tops and between the decay products of the $W$-bosons reduce this to $6!/(2 \times 2 \times 2) = 90$ permutations. Further complicating things, ∼50% of $t\bar{t}$ events at the LHC are expected to contain at least one additional jet which is not the result of top quark decay, leading to $7!/(2 \times 2 \times 2) = 630$ or $8!/(2 \times 2 \times 2 \times 2) = 2520$ permutations[1] in 7- or 8-jet events, respectively. Higher jet multiplicity events are also likely; nonetheless, the current state-of-the-art techniques depend on enumerating and evaluating each permutation to identify the best candidate. The many incorrect assignments obscure the true assignment, diluting the scientific power of the data, and represents a significant computational penalty when all permutations are evaluated for every event. Sometimes, each permutation must be evaluated *per systematic uncertainty* per event, which is often intractable in the datasets typical in high-energy-physics (HEP).

The most common technique is a $\chi^2$-minimization method, which scores a permutation based on the consistency of the reconstructed $W$-boson masses with known values and similarity of the two reconstructed top quark masses[2]:

$$\chi^2 = \frac{(m_{bqq} - m_{b'q'q'})^2}{\sigma^2_{\Delta m_{bqq}}} + \frac{(m_{qq} - m_W)^2}{\sigma^2_W} + \frac{(m_{q'q'} - m_W)^2}{\sigma^2_W}$$

(1)

We set $m_W = 81.3$ GeV, and find for our dataset (described in Sec. IV) values of $\sigma_W = 12.3$ GeV and $\sigma_{\Delta m_{bjj}} = 26.3$ GeV following a Gaussian fit to the relevant distributions. $\chi^2$ is evaluated for every permutation, and the parton assignment with the minimum value is chosen. This method typically uses $b$-tagging to consider only permutations with $b$-jets in the place of $b$-quarks. This reduces the permutations but prevents the correct solution being found in the presence of mistagged jets. We apply this requirement in our studies for consistency with recent experimental results—this implementation has been the preferred reconstruction for ATLAS [1,10], while CMS have used a similar method [11].

---

[1]The additional factor of 2 is the invariance of the extraneous jets.
[2]Explicitly using $m_{top}$ in the $\chi^2$ equation improves reconstruction efficiency by around 5% at the expense of dramatically sculpting the top mass distribution. We chose to make comparisons to the definition used in recent experimental results.

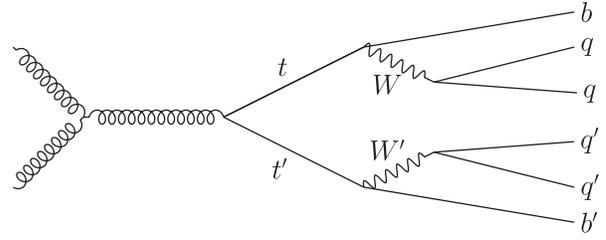

FIG. 1. Production diagram of a top quark pair, $tt'$, and their decays into $qqb$ and $q'q'b'$, respectively.

Another approach is the use of boosted decision trees or neural networks as permutation classifiers, though these have mostly been employed in the leptonic channels where combinatoric explosion is reduced [12,13], or to reconstruct tops individually [14]. CMS has also used a hybrid $\chi^2$ and BDT method in the all-jet channel [15].

A more advanced minimization technique is implemented in KLFitter [16]. Transfer functions are used to represent the detector response in a likelihood function per permutation, which is minimized to find the assignments. This operates similarly to the $\chi^2$ technique, though has greater CPU requirements due to the more complex model. KLFitter has to date been almost exclusively used in the single lepton channel, and with the $\chi^2$ already CPU limited in many cases, it is not discussed here.

## III. INVARIANCE AND ATTENTION IN DEEP NEURAL NETWORKS

Equivariance and invariance properties can play an important role in the design of both feed-forward and recursive neural networks across different forms of learning [17–22]. For instance, classical convolution neural networks can produce object recognition outputs that are invariant with respect to translations in their two-dimensional inputs, and this invariance property has been generalized to apply to other manifolds and groups [23,24]. In the problem considered here, the network output should be invariant under permutations of the input jet order. Such permutation invariance has been explored in set-based [25–27] and graph networks [28,29]. The output should further identify two distinct interchangeable triplets, $qqb$ and $q'q'b'$, each including an interchangeable pair, $qq$ or $q'q'$. This permutation invariance on the output is a unique property of our dataset which our architecture must account for.

Attention mechanisms allow the network to selectively propagate information (gating)—the activities of a set of neurons are multiplied, component-wise, by the activities of another set of neurons. These gating mechanisms allow neural networks to dynamically modulate neuron activity as a function of the other neurons or inputs. Recursive and attention-based networks, which allow the network to infer relationships between different elements in a sequence, have achieved state-of-the-art performance in a natural language processing in machine translation [30], language understanding [31], and text generation [32].

Attention architectures are permutation invariant because rearranging the order of the elements in the input sequence induces the same rearrangement in the attention weights. This feature can be leveraged to endow the network with permutation symmetry [25,26]. We leverage the permutation invariance present in attention-based methods to efficiently model the symmetries of the top quark pair system. We generalize the ideas present in dot-product attention to allow for a three-way symmetry-preserving attention mechanism which can perform jet assignments into $qqb$ and $q'q'b'$ triplets.

## IV. DATASETS

A sample of 50M simulated $pp \rightarrow t\bar{t}$ events was generated at $\sqrt{s} = 13$ TeV using MADGRAPH_AMC@NLO [33] (v2.7.2), interfaced to PYTHIA8 [34] (v8.2) for showering and hadronization. Detector response was simulated using DELPHES [35] (v3.4.2) with the ATLAS parametrization. Events are generated at leading order in quantum-chromodynamics (QCD), with the top mass $m_{\text{top}} = 173$ GeV, and the $W$-boson forced to decay hadronically. Jets, reconstructed using the anti-$k_T$ algorithm [36] as implemented in FASTJET [37] (v3.2.1) with radius $R = 0.4$, are required to have transverse momentum $p_T \geq 25$ GeV and absolute pseudorapidity $|\eta| < 2.5$, and are tagged as originating from $b$-quarks with $p_T$-dependent efficiency and mistag rates. 5,926,407 events meet the preselection requirements of $\geq 6$ jets and $\geq 2$ $b$-jets. This dataset is available at [38].

The supervised learning technique employed here requires a training sample in which the correct assignments are identified. We define the correct jet assignments by matching them to the simulated truth quarks within an angular distance of $\Delta R = \sqrt{\Delta\eta^2 + \Delta\phi^2} < 0.4$. Requiring all six quarks to be unambiguously matched leaves 2,967,955 total events for training, and 119,283 for performance evaluation and hyperparameter tuning. This matching requirement has an efficiency of 24% on 6-jet events, 32% on 7-jet events, and 40% on $\geq 8$-jet events. We also evaluate the performance on 365,939 additional testing events which contain any number of reconstructable top quarks. Of these events, 191,597 contain only one top quark with all decay products matched, with filter efficiencies of 56%, 52%, and 48% in 6-jet, 7-jet, and $\geq 8$-jet events respectively.

We note that the matching of partons to jets via the parton-shower history in this way is not strictly physical in QCD, as these partons are not directly observable in a physical system and are modified by color correlations with the rest of the event which may not have been fully taken into account at the level of the matching procedure [39]. Nonetheless, this is a common paradigm, including in the

$\chi^2$ benchmark process [1,10], which we adopt in order to define the learning task. The physical interpretation of any measurement which utilizes our technique is unaffected by this procedure, as the output of the network is used only to pick the jet triplets which are to be measured. This is exactly analogous to the use of machine-learning based boosted top taggers [40,41].

## V. SPA-NET ARCHITECTURE

We perform the jet classification with SPA-NET: an attention-based neural network which encodes the symmetries of the problem. Inputs to the network are an unsorted list of jets, each represented by their 4-vector ($p_T, \eta, \phi, M$) as well as a boolean $b$-tag. $M$ and $p_T$ are logarithmically scaled, and then each component is independently normalized to have zero mean and unit variance. The network, shown in Fig. 2, consists of six components: a jet-independent embedding which converts each jet into a $D$-dimensional latent space representation; a stack of transformer encoders which learn contextual relationships; two additional transformer encoders on each branch to extract top-quark information; and two tensor-attention layers to produce the top-quark distributions. The transformer encoders employ a variant of attention known as multi-head self-attention [30], though they may use any permutation-invariant architecture in general.

### A. Symmetry preserving tensor attention

Jet assignment in the context of $t\bar{t}$ events presents several unique problems for typical classification networks, as a variety of symmetries complicate the output generation and training. Primarily, the physics quantities are invariant under permutation of the $W$ decay products $qq$. The network must not differentiate between predictions which prefer one ordering to the other, and match each $qq$ pair to the appropriate $b$. This naturally creates a triplet relation $qqb$, with $qq$ obeying permutation symmetry. In order to encode this into the network, we develop a generalization of the attention mechanism which can learn $n$-way relationships with included symmetries. We call this technique *tensor attention*.

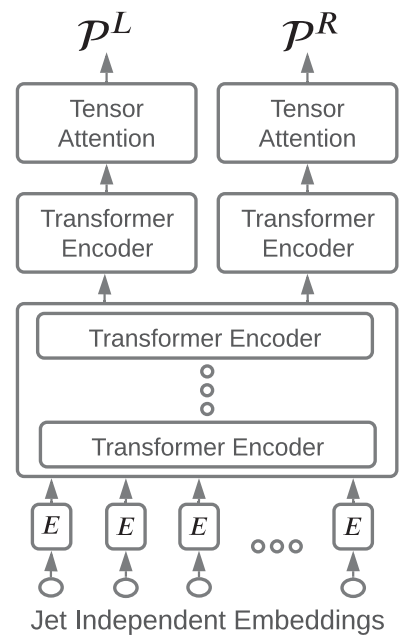

FIG. 2. High-level structure of SPA-NET.

Each tensor attention layer contains a set of weights $\theta \in \mathbb{R}^{D \times D \times D}$. This tensor is not inherently symmetric: in order to produce an invariant attention weighting, we first transform it into an auxiliary weights tensor which conforms to the classification permutation group. We produce $S \in \mathbb{R}^{D \times D \times D}$ and use it to perform weighted dot-product attention on a list of embedded jets $X \in \mathbb{R}^{N \times D}$, where $N$ is the number of jets. Working in flat Euclidean space, we express the attention mechanism in Einstein notation, with

$$S^{ijk} = \frac{1}{2}(\theta^{ijk} + \theta^{jik}) \qquad (2)$$

$$O^{ijk} = X_n^i X_m^j X_l^k S^{nml}. \qquad (3)$$

The summation in Eq. (2) guarantees that the first two dimensions of $S$ will be symmetric, ensuring that $O^{ijk} = O^{jik}$, and enforcing $qq$ invariance. Afterwards, we perform a 3-dimensional softmax on $O$ to generate the joint triplet probability distribution

$$\mathcal{P}(i, j, k) = \frac{\exp O^{ijk}}{\sum_{i,j,k} \exp O^{ijk}} \qquad (4)$$

We produce individual distributions for each of the two top-quarks, $\mathcal{P}^L$ and $\mathcal{P}^R$, and we produce a single triplet from each by selecting the peak of these distributions.

An important note is that the weights tensor rank depends only on the hyperparameter $D$, and thus, it is possible to include any number of jets in each event. Furthermore, the network evaluation scales only as $\mathcal{O}(N^3)$ with respect to the number of jets because we only need to produce a triplet distribution $\mathcal{P}$. This removes a crippling limitation of the $\chi^2$ method, which grows as $\mathcal{O}(N^6)$. In our dataset, the largest jet multiplicity in a single event is $N = 18$.

### B. Training

We train these distributions via cross-entropy between the output probabilities and the true target distribution on the all-jet $t\bar{t}$ problem, naming the resulting network SPA$t\bar{t}$ER (SPA-NET for $t\bar{t}$ reconstruction). This formulation contains another symmetry which can be exploited: the top quark pairs are invariant with respect to the labels $tt' \leftrightarrow t't$. We create a symmetric loss function based on cross-entropy, $H(X, Y) = \sum_{(x,y) \in (X,Y)} -x \log(y)$, which allows either of the networks two output distributions, $\mathcal{P}^L$ and $\mathcal{P}^R$, to match either one of the targets $\mathcal{T}_1$ and $\mathcal{T}_2$. The target distributions $\mathcal{T}$ have two symmetric nonzero entries, one for each permutation of the $qq$ pair. The loss $\mathcal{L}$ is expressed as

$$\mathcal{L} = \min \left( \mathcal{L}_1(\mathcal{P}^L, \mathcal{T}_1, \mathcal{P}^R, \mathcal{T}_2), \mathcal{L}_1(\mathcal{P}^L, \mathcal{T}_2, \mathcal{P}^R, \mathcal{T}_1) \right) \qquad (5)$$

where $\mathcal{L}_1(P_1, T_1, P_2, T_2) = H(T_1, P_1) + H(T_2, P_2)$. The resulting distributions may classify the same jet to be part of both triplets. To enforce unique predictions, we select the assignment of the higher probability $\mathcal{P}$ first, and set the colliding bins of the other $\mathcal{P}$ to zero. We then evaluate the second $\mathcal{P}$ to select the best noncontradictory classifications.

We also note that SPA$t\bar{t}$ER does not enforce that $b$-tagged jets are selected in the position of the $b$-quarks. This allows the network to correctly predict events in which there are mistagged jets, while still utilizing $b$-tagging information. In the $\chi^2$, allowing this means $b$-tagging information is completely lost, and greatly increases the number of permutations.

SPA$t\bar{t}$ER contains 2.1M parameters in each tensor attention layer and 600 k parameters in the central transformer encoder stack. SPA$t\bar{t}$ER was trained using the ADAMW optimizer [42] and four Nvidia Titan-V GPUs, converging after approximately 4 hours.

## VI. PERFORMANCE

To assess performance we define two metrics, which can only be evaluated on *identifiable* top quarks where the correct assignment has been identified as described previously. The first metric is $\epsilon^{\text{top}}$, the fraction of identifiable top quarks which have all three jets correctly assigned. This is reported in two event subsets, those where only one top quark is identifiable ($\epsilon_1^{\text{top}}$), and those where both top quarks are identifiable ($\epsilon_2^{\text{top}}$). We further define $\epsilon^{\text{event}}$, the fraction of events with two identifiable top quarks in which both top quarks have all jets correctly assigned. Table I shows these metrics for both the $\chi^2$ and SPA$t\bar{t}$ER methods.

The $\chi^2$ has an $\epsilon^{\text{event}}$ of 41.2%, while SPA$t\bar{t}$ER achieves an $\epsilon^{\text{event}}$ of 63.7%. The $\chi^2$ suffers from a large reduction in performance as jet multiplicity increases, peaking at 55.2% for events with exactly 6 jets and dropping to 20.5% in events with at least 8 jets. This drop is less pronounced for SPA$t\bar{t}$ER, at 80.7% in 6-jet events and 52.3% in ≥8-jet events. We observe a similar trend in the per-top efficiencies; for $\epsilon_2^{\text{top}}$, SPA$t\bar{t}$ER achieves 73.5% inclusively, compared to just 49.7% for the $\chi^2$. These numbers drop to 66.2% and 33.6% respectively in ≥8-jet events. The performance is lower in events in which only one top is identifiable, though SPA$t\bar{t}$ER still strongly outperforms the $\chi^2$, with an $\epsilon_1^{\text{top}}$ of 55.2% and 28.6% respectively. We also note that in our evaluation dataset, 8.1% of events in which both tops are identifiable have at

 

TABLE I. Performance of the $\chi^2$ and SPA$t\bar{t}$ER assignments assessed by per-event efficiency $\epsilon^{\text{event}}$ and per-top efficiencies $\epsilon^{\text{top}}$, inclusively and by jet multiplicity $N_{\text{jets}}$.

| $N_{\text{jets}}$ | $\chi^2$ Method | | | SPA$t\bar{t}$ER | | |
|---|---|---|---|---|---|---|
| | $\epsilon^{\text{event}}$ | $\epsilon^{\text{top}}_2$ | $\epsilon^{\text{top}}_1$ | $\epsilon^{\text{event}}$ | $\epsilon^{\text{top}}_2$ | $\epsilon^{\text{top}}_1$ |
| 6 | 55.2% | 59.6% | 28.9% | 80.7% | 84.1% | 56.7% |
| 7 | 36.6% | 47.4% | 29.8% | 66.8% | 75.7% | 56.2% |
| ≥8 | 20.5% | 33.6% | 25.5% | 52.3% | 66.2% | 52.9% |
| *Inclusive* | **41.2%** | **49.7%** | **28.6%** | **63.7%** | **73.5%** | **55.2%** |

least one $b$-quark matched to non-$b$-tagged jets. These quarks, which are impossible for the $\chi^2$ to correctly reconstruct, are reconstructed by SPA$t\bar{t}$ER with an efficiency of 29.4%.

We inspect the reconstructed $W$ mass using the assignments generated by both methods in Fig. 3, broken down into three categories: "correct," "incorrect," and "unmatched," corresponding to the cases in which: all three top decay products are correctly assigned, all three top decay products are present in the event but at least one is incorrectly assigned, and one or more of the top decay products is not identifiable, respectively. The $\chi^2$ has a narrower peak around $m_W$ than SPA$t\bar{t}$ER, though much of this shape comes from the incorrect and unmatched events. This is explained by the presence of $m_W$ in Eq. (1), and demonstrates that SPA$t\bar{t}$ER is utilizing more information than just the invariant mass peak. Figure 4 shows the $m_{\text{top}}$ distributions for the same three categories. SPA$t\bar{t}$ER has the more peaked distribution, with the $\chi^2$ showing a much larger tail to high masses in the incorrect and unmatched events. The correctly reconstructed events are centred at the expected masses, and the fraction of events in this category is larger with SPA-NET than the $\chi^2$, as expected. Additional distributions used for cross-checks are available in Appendix A (Figs. 5–12).

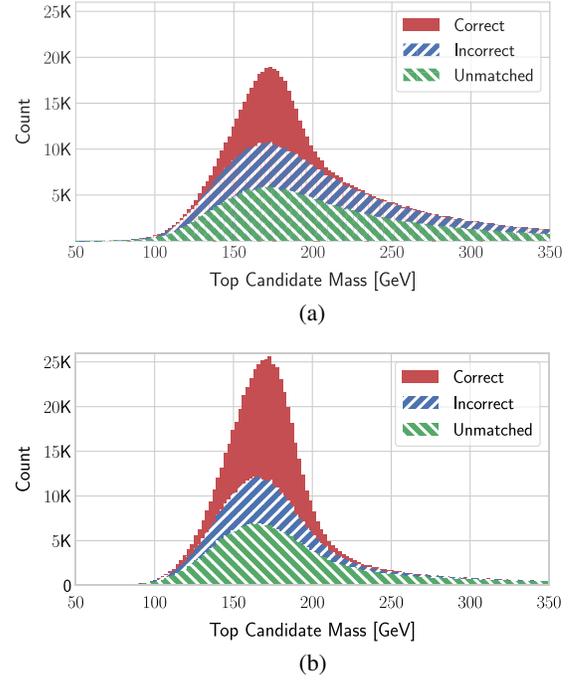

FIG. 4. Stacked distributions of reconstructed $m_{\text{top}}$ using (a) the $\chi^2$, and (b) SPA$t\bar{t}$ER.

In 11.2% of fully matched events, the network predicted the same jet to be part of both top quarks. SPA$t\bar{t}$ER correctly predicts 25.0% of these events after the reselection described above, compared to only 12.7% for the $\chi^2$, indicating that these were generally difficult events to classify correctly. We also note that the average softmax value of the predicted jets in these events is only 36%, compared to 74% for all fully reconstructable events.

A final important performance metric is the computation time per event. The time required to evaluate the $\chi^2$ scales approximately as $P(N, 6) = \mathcal{O}(N^6)$ with the number of jets in the event, and this often leads to analyses setting a maximum number of jets to consider, degrading the performance for purely CPU-time reasons. On the author's laptop, a 2019 Dell XPS13 with an Intel-Core i7-1065G7 1.30 GHz CPU, SPA$t\bar{t}$ER took an average of 4.4 ms to evaluate per event, with no dependence on jet multiplicity.[3] A further order of magnitude speed improvement is found when evaluating on a GPU. In contrast, the $\chi^2$ took an average of 20 ms in 6-jet events, 48 ms in 7-jet events, and 369 ms in ≥8-jet events.

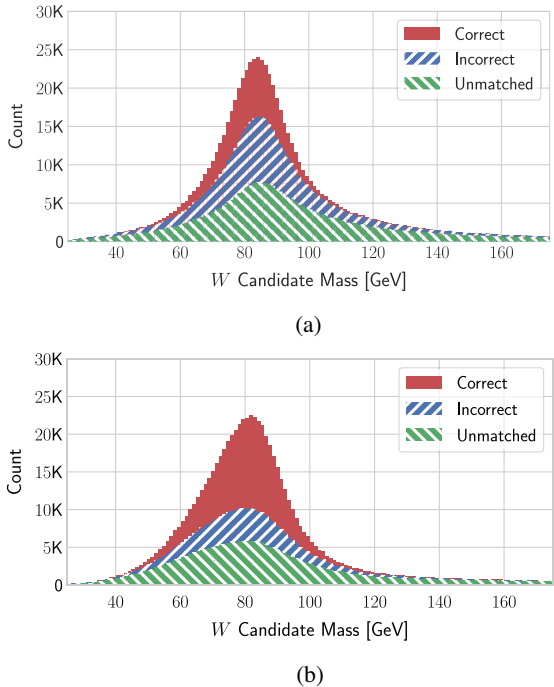

FIG. 3. Stacked distributions of reconstructed $m_W$ using (a) the $\chi^2$, and (b) SPA$t\bar{t}$ER.

---

[3]We pad all events to the maximum of 18 jets, making evaluation time constant vs $N_{\text{jets}}$.

## VII. CONCLUSIONS

SPA-NET's have inherent permutation symmetries which make them very well suited to the task of jet assignment, where the permutations otherwise lead to an increase in computation time and dilutes the scientific value of the data. Our network SPA*t̄t*ER demonstrates superior performance on this task. The adoption of our technique by the experimental collaborations ATLAS and CMS will lead to significantly improved precision of analyses in the all-jet *t̄t* final-state by improving the fraction of events that are well reconstructed from 41.2% using existing methods to 63.7% using our new technique. This paradigm shift will allow greatly enhanced sensitivity in high jet multiplicity events making many new analyses viable in this final-state.

This paper describes just one of many possible applications of SPA-NET's to event reconstruction in HEP. Future work may include extending these techniques to alternative all-jet final-states, to other *t̄t* decay modes, or many other classes of problem. Though not studied here, the output of SPA*t̄t*ER may also be used in additional ways, such as setting a minimum reconstruction quality requirement that will act to suppress backgrounds, analogously to how the $\chi^2$ is used in [1]. Additional input information, such as jet substructure [43] or (pseudo-)continuous *b*-tagging [44], may also improve performance.

This letter contributes to a family of work which help endow machine learning methods with problem specific invariances. We have presented an efficient polynomial-time approach for tackling classification tasks where the targets must obey a set of permutation symmetries. Such symmetries underlie the mathematical foundation of the Standard Model, but they may be found in many other common classification tasks, such as graph matching [45] and hierarchical clustering algorithms. Well trained deep neural networks can replace permutation based algorithms, and avoid combinatorial explosion, by effectively estimating symmetry-aware pair-wise similarities. Symmetries can also be used to create smaller models that reduce the amount of data necessary for training [46]. Understanding and exploiting the invariances present in any modeling task is vital for effective learning.

Our code is available at [38].

## ACKNOWLEDGMENTS

M. F would like to thank Nicole Hartman for useful early discussions, and Megan Remillard for language editing assistance. D. W. and M. F. are supported by the U.S. Department of Energy (DOE), Office of Science under Grant No. DE-SC0009920. S.-C. H. is supported by the U.S. Department of Energy, Office of Science, Office of Early Career Research Program under Award No. DE-SC0015971. The work of A. S. and P. B. in part supported by Grants No. NSF NRT 1633631 and No. ARO 76649-CS to P. B. The work of T.-W. H. was supported by the Taiwan MoST with the Grant No. MOST-107-2112- M-007-029-MY3. M. F. and A. S. contributed equally to this work.

## APPENDIX A: SUPPLEMENTARY PLOTS

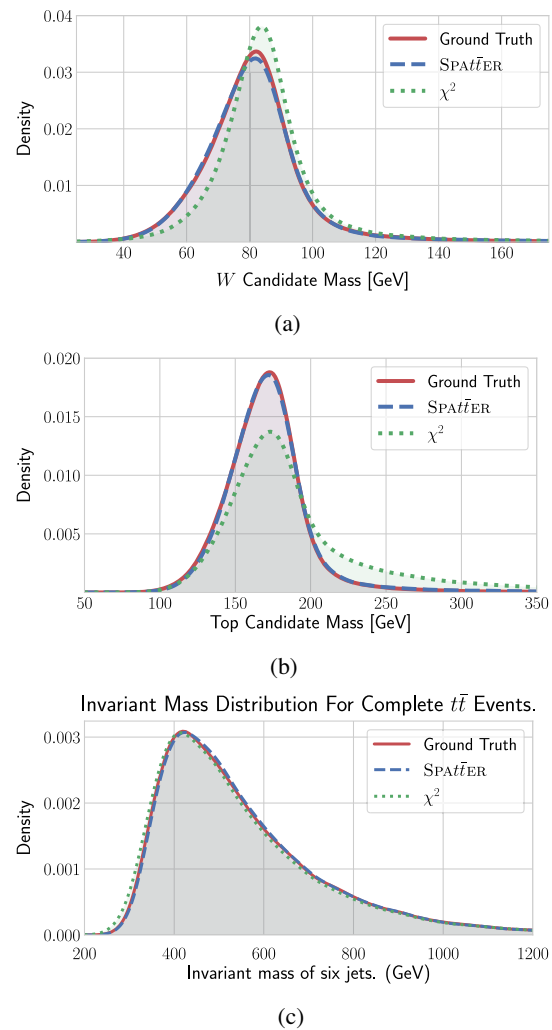

FIG. 5. Comparison of (a) $m_W$, and (b) $m_{\text{top}}$, for the truth matched jets (solid red), the $\chi^2$ assignments (dotted green), and the SPA*t̄t*ER assignments (dashed blue). A Gaussian KDE has been applied.

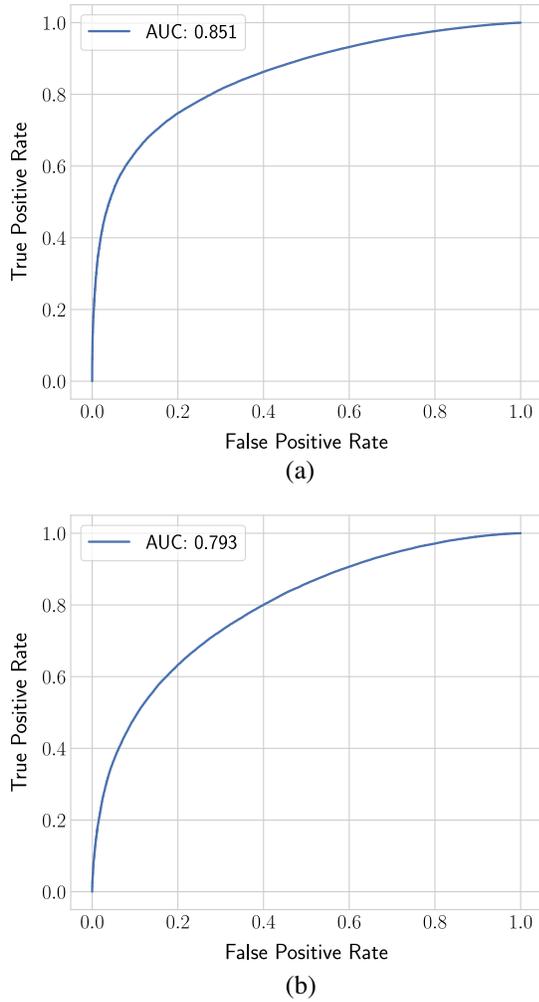

**FIG. 6.** Receiver operating characteristic (ROC) curve of the predicted top-quark triplet softmax value for SPAɪ̃TER on events with (a) two and (b) one reconstructable tops. Targets are defined as 1 if the predicted triplet was correct and 0 and otherwise. Included is the area under each curve.

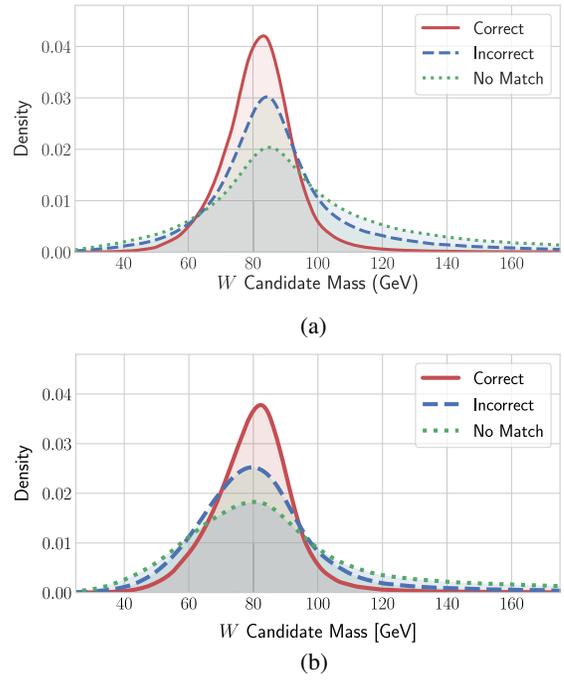

**FIG. 7.** Normalized distributions of reconstructed $m_W$ using (a) the $\chi^2$, and (b) SPAɪ̃TER.

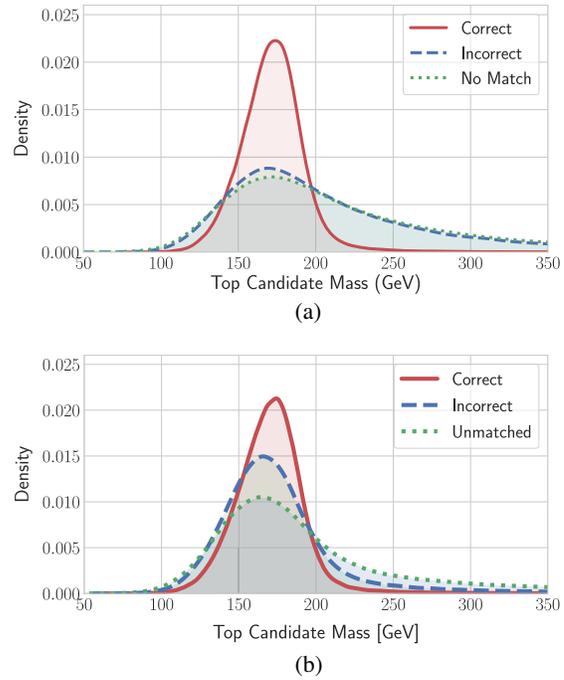

**FIG. 8.** Normalized distributions of reconstructed $m_{\text{top}}$ using (a) the $\chi^2$, and (b) SPAɪ̃TER.

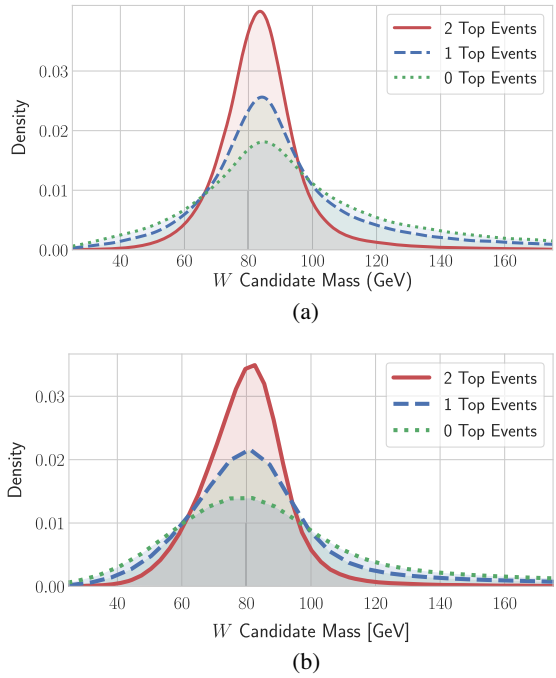

FIG. 9. Normalized distributions of reconstructed $m_{\text{top}}$ using (a) the $\chi^2$, and (b) SPAɴ*t̄t*ER.

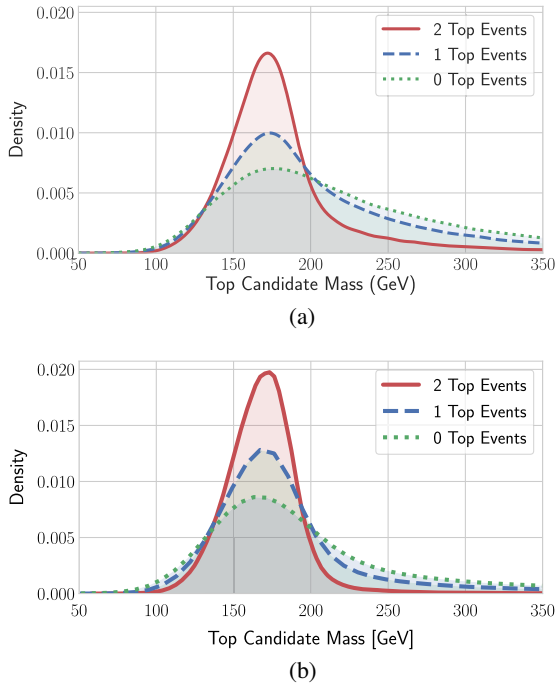

FIG. 10. Normalized distributions of reconstructed $m_{\text{top}}$ using (a) the $\chi^2$, and (b) SPAɴ*t̄t*ER.

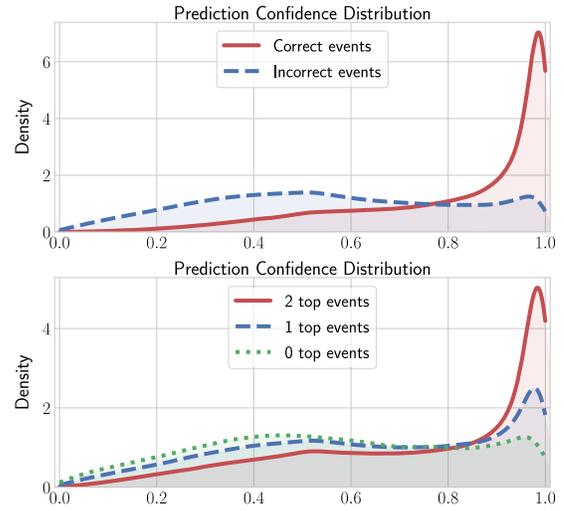

FIG. 11. The distribution of the predicted triplet's softmax value in SPAɴ*t̄t*ER's output distributions, grouped by either correctness or event type.

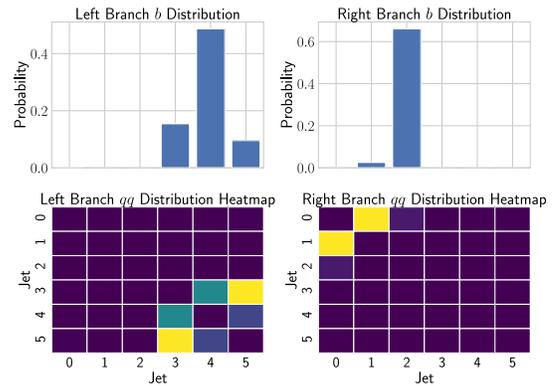

FIG. 12. A visualization of an example output for a single event. The top plots display the projected $b$ distribution, and the lower plots show the joint $qq$ distributions.

## APPENDIX B: SPAɴ*t̄t*ER MATHEMATICAL FORMULATION

In this section, we provide a more detailed description of SPAɴ*t̄t*ER's architecture which is necessary to recreate the model. Figure 2 provides a high-level graphical overview of the network. Here we describe the mathematics necessary to implement each of the sections in the figure. SPAɴ*t̄t*ER has a very recursive architecture, so we define several terms to simplify this description. SPAɴ*t̄t*ER consists of several large *stacks*, which are compositions of several identically structured *blocks*, each of which contains one or more *layers*, and each of which contains one or more *parameters*.

## 1. Independent embedding

The *embedding stack* consists of several embedding blocks, each of which progressively increases the latent dimensionality of the input jets up to the final dimensionality $D$. The *embedding blocks* are feed-forward, fully connected neural networks which are applied independently to each jet via weight sharing. Each block consists of a fully connected matrix multiplication layer $L$ with parameters $W \in \mathbb{R}^{D_o \times D_i}$ and $b \in \mathbb{R}^{D_o}$; a *PReLU* [47] nonlinearity with a parameter $a \in \mathbb{R}^{D_o}$; and a one-dimensional *BatchNorm* [48] layer with parameters $\mu, \sigma \in \mathbb{R}^{D_o}$. A single embedding block with an output dimensionality $D_o$ can be described as $E_{D_o} = BatchNorm \circ PReLU \circ L$ where

$$BatchNorm(x) = \frac{x - E[x]}{\sqrt{\text{Var}[x]}} * \sigma + \mu$$

$$PReLU(x) = \max(x, 0) + a * \min(x, 0)$$

$$L(x) = Wx + b$$

We stack several of these embedding block, with each block doubling their latent space dimensionality, starting from the original 5 jet features. SPAɴᴇᴛ has a target latent space dimensionality of $D = 128$, so the $D_o$ values follow the sequence: $8 \rightarrow 16 \rightarrow 32 \rightarrow 64 \rightarrow 128$. The full embedding stack can be described as the following composition:

$$E = E_{128} \circ E_{64} \circ E_{32} \circ E_{16} \circ E_8 \tag{B1}$$

## 2. Transformer encoder

The *encoder stack* consists of a sequence of transformer encoder blocks as described by Vaswani *et al.* [30]. A single *encoder block* contains a multi-head attention layer $Attention(Q, K, V)$ [30]; two $LayerNorm$ operations [49]; two feed-forward layers $L_1$ and $L_2$; and a $PReLU$ nonlinearity. In this paper, we provide a general overview of the encoder structure, but we omit a detailed description of the multihead attention and layer normalization operations. A single transformer encoder block $T_i$ can be expressed as $T_i = B \circ A$, where

$$B(x) = LayerNorm(Attention(x, x, x) + x) \tag{B2}$$

$$A(x) = LayerNorm(L_2(PReLU(L_1(x))) + x) \tag{B3}$$

The complete transformer encoder stack, $T$, is simply a composition of $k$ encoder blocks, one after the other. We use $k = 6$ in SPAɴᴇᴛ.

$$T = T_k \circ T_{k-1} \circ \ldots \circ T_2 \circ T_1 \tag{B4}$$

## 3. Branch encoders

After passing through the shared, central embedding and encoder stacks, SPAɴᴇᴛ's signal path splits into two branches, one for each of the target top quarks. Figure 2 demonstrates this branch splitting. We name these two paths the left and right branch, although this distinction is arbitrary. Each of these branches contains an independent embedding and encoder stack. The *branch embedding stacks* share a near-identical structure to the initial independent embedding stack, except they preserve the latent dimensionality: $D_i = D_o$. The embedding blocks $E_i^L$ and $E_i^R$ are applied independently to each jet. The *branch encoder stacks* also share an identical structure to the central encoder stack. The left branch, $T^L$, and right branch, $T^R$, can be described as compositions of $j$ embedding blocks and $l$ transformer encoder block. We use $j = l = 4$ in SPAɴᴇᴛ.

$$T^L = T_l^L \circ \ldots \circ T_1^L \circ E_j^L \circ \ldots \circ E_1^L \tag{B5}$$

$$T^R = T_l^R \circ \ldots \circ T_1^R \circ E_j^R \circ \ldots \circ E_1^R \tag{B6}$$

## 4. Tensor attention

We provide a detailed description of the tensor attention output layers in the main text with Eqs. (2)–(4). Here, we replicate the description in a concise manner.

Each tensor attention layers contains a single parameter $\theta \in \mathbb{R}^{D \times D \times D}$. We produce an intermediate symmetric tensor $S \in \mathbb{R}^{D \times D \times D}$ who's indices obey the symmetry group of the top quark triplet. This is combined with the list of input vectors $X \in \mathbb{R}^{N \times D}$ into an output tensor $O \in \mathbb{R}^{N \times N \times N}$. Finally, this output tensor is passed through a softmax nonlinearity in order to produce valid three-way joint distributions $\mathcal{P}^L$ and $\mathcal{P}^R$. The complete equations for this layer are

$$S^{ijk} = \frac{1}{2}(\theta^{ijk} + \theta^{jik}) \tag{B7}$$

$$O^{ijk} = X_n^i X_m^j X_l^k S^{nml} \tag{B8}$$

$$\mathcal{P}(i, j, k) = \frac{\exp O^{ijk}}{\sum_{i,j,k} \exp O^{ijk}} \tag{B9}$$

## 5. Efficient tensor attention

Equation (B8) can be expression as a cubic tensor form, which naively requires $\mathcal{O}(N^3 D^3)$ operations. When actually computing this expression, we have to construct several very large intermediate tensors. Since we use optimized GPU linear algebra libraries in order to perform tensor operations, we have to split the evaluation into several, more fundamental, operations. We use opt_einsum [50] in order to generate an optimal set of operations for Eq. (B8). The summation can be expression using the following intermediate operations, each of which is a generalized matrix-multiplication (GEMM).

$$A^{mli} = S^{nml} X_n^i$$

$$B^{lij} = A^{mli} X_m^j$$

$$O^{ijk} = B^{lij} X_l^k$$

With the intermediate tensors $A^{mli}$ and $B^{lij}$ have dimensionalities $(D \times D \times N)$ and $(D \times N \times N)$ respectively. Since $D > N$ in our situation, this operation requires a large amount of memory to store all of these intermediate components.

Instead of storing the complete cubic weights tensor $\theta$ and explicitly finding the summation, we limit the possible $\theta$ weights that we can learn in order to greatly reduce the intermediate tensors. Instead of learning $\theta \in \mathbb{R}^{D \times D \times D}$, we decompose $\theta$ into three matrices $\theta_1, \theta_2, \theta_3 \in \mathbb{R}^{D \times D}$. Then we compute three intermediate vectors from $X$ by simply performing regular matrix multiplication

$$A = \theta_1 X$$

$$B = \theta_2 X$$

$$C = \theta_3 X$$

These can be easily implemented with fully connected neural network layers and computed in parallel. Finally, we can estimate our original output by performing a trivial three-form with these tensors.

$$O^{ijk} = A_n^i B_m^j C_l^k \mathbb{1}^{nml}$$

where $\mathbb{1}^{nml} = 1$ for all possible indices. The space complexity of this decomposition when including the intermediate tensors reduces to $\mathcal{O}(N^3 + ND + D^2) \approx \mathcal{O}(D^2)$, and improvement over the naive approach which has space complexity of $\mathcal{O}(N^3 + ND^2 + D^3) \approx \mathcal{O}(D^3)$. Additionally, when using opt_einsum to evaluate the three-form, the runtime is reduced from $\mathcal{O}(N^3 D^3)$ to $\mathcal{O}(N^3 D^2)$.

We can replicate the symmetry constrain we impose in Eq. (B7) by simply requiring the first two decomposed matrices to be equal to each other.

$$\theta_1 = \theta_2$$

This decomposition is not one-to-one, so not every $\theta$ can be represented in this form. However, in our experiments, we found no drop in predictive performance when using this decomposition.

## 6. Batching

Since most deep neural network implementations use efficient batch matrix multiplication routines in order speed up computation, it is beneficial to feed batches of several events into the network simultaneously. However, events can contain a varying number of jets: a single event can have a length of anywhere from 6 to 20 momentum vectors. In order to process all of these events together in batches, we pad all events to the maximum size of 20 jets by appending 0-vectors, creating a batched input tensor $X \in \mathbb{R}^{B \times 20 \times 5}$ where $B$ is the batch size. We also construct a secondary *masking* input $M \in \{0, 1\}^{B \times 20}$. This vector indicates if the given jet in a given event is a real jet or a padding jet. This vector is employed during multi-head attention in order to prevent the attention weights from including the masked jets [30]. The masking vector is also employed during softmax calculation to prevent the masked vectors from skewing the output distributions. When applied to a batch output tensor $O \in \mathbb{R}^{B \times 20 \times 20 \times 20}$, the softmax calculation can be described as:

$$\mathcal{P}(b, i, j, k) = \frac{M^{bi} M^{bj} M^{bk} \exp O^{bijk}}{\sum_{i,j,k} M^{bi} M^{bj} M^{bk} \exp O^{bijk}} \quad \text{(B10)}$$

## 7. Hyperparameters

We select optimal hyperparameters for SPAɴɪTER by running a parallel Gaussian process hyperparameter search using the SHERPA hyperparameter optimization library [51]. We evaluate 500 different sets of hyperparemeters using a subset of the original training dataset, training on only the first 50% of events. We also sample the last 5% of the training dataset to act as the validation dataset for the hyperparameter search. We evaluate each model by computing the event purity, $\epsilon_2^{\text{top}}$, on this validation dataset. The first 100 sets of hyperparameters are randomly sampled from a uniform distribution. Afterwards, we use a Gaussian process optimizer in order to suggest future hyperparamers which may improve the purity. A complete listing of the final hyperparameters for this network can be found in Table II.

TABLE II. A summary of hyperparameter values for SPAɴɪTER, decided via hyperparameter optimization.

| Name | Symbol | Value |
|---|---|---|
| Latent Dimensionality | $D$ | 128 |
| Encoder Count | $k$ | 6 |
| Branch Embedding Count | $j$ | 5 |
| Branch Encoder Count | $l$ | 3 |
| Batch Size | $B$ | 4096 |
| Learning Rate | $\alpha$ | $1.5 \times 10^{-3}$ |

[1] ATLAS Collaboration, Measurements of top-quark pair single- and double-differential cross-sections in the all-hadronic channel in $pp$ collisions at $\sqrt{s} = 13$ TeV using the ATLAS detector, J. High Energy Phys. 01 (2021) 033.

[2] CMS Collaboration, Measurement of differential $t\bar{t}$ production cross sections using top quarks at large transverse momenta in pp collisions at $\sqrt{s} = 13$ TeV, Phys. Rev. D **103**, 052008 (2021).

[3] ATLAS Collaboration, Search for the Standard Model Higgs boson decaying into $b\bar{b}$ produced in association with top quarks decaying hadronically in pp collisions at $\sqrt{s} = 8$ TeV with the ATLAS detector, J. High Energy Phys. 05 (2016) 160.

[4] ATLAS Collaboration, Search for resonances decaying into top-quark pairs using fully hadronic decays in $pp$ collisions with ATLAS at $\sqrt{s} = 7$ TeV, J. High Energy Phys. 01 (2013) 116.

[5] CMS Collaboration, Search for vector-like T quarks decaying to top quarks and Higgs bosons in the all-hadronic channel using jet substructure, J. High Energy Phys. 06 (2015) 080.

[6] A. Abdesselam et al., Boosted objects: A probe of beyond the standard model physics, Eur. Phys. J. C **71**, 1661 (2011).

[7] ATLAS Collaboration, Measurements of top-quark pair differential and double-differential cross-sections in the $\ell$ + jets channel with $pp$ collisions at $\sqrt{s} = 13$ TeV using the ATLAS detector, Eur. Phys. J. C **79**, 1028 (2019); Erratum, Eur. Phys. J. C **80**, 1092 (2020).

[8] P. Zyla et al. (Particle Data Group), Review of particle physics, Prog. Theor. Exp. Phys. **2020**, 083C01 (2020).

[9] ATLAS, CDF, CMS, and D0 Collaborations, First combination of Tevatron and LHC measurements of the top-quark mass, arXiv:1403.4427.

[10] ATLAS Collaboration, Top-quark mass measurement in the all-hadronic $t\bar{t}$ decay channel at $\sqrt{s} = 8$ TeV with the ATLAS detector, J. High Energy Phys. 09 (2017) 118.

[11] CMS Collaboration, Measurement of the top quark mass in the all-jets final state at $\sqrt{s} = 13$ TeV and combination with the lepton + jets channel, Eur. Phys. J. C **79**, 313 (2019).

[12] J. Erdmann, T. Kallage, K. Krninger, and O. Nackenhorst, From the bottom to the top reconstruction of $t\bar{t}$ events with deep learning, J. Instrum. **14**, P11015 (2019).

[13] ATLAS Collaboration, Search for the standard model Higgs boson produced in association with top quarks and decaying into a $b\bar{b}$ pair in $pp$ collisions at $\sqrt{s} = 13$ TeV with the ATLAS detector, Phys. Rev. D **97**, 072016 (2018).

[14] ATLAS Collaboration, $CP$ Properties of Higgs Boson Interactions with Top Quarks in the $t\bar{t}H$ and $tH$ Processes Using $H \to \gamma\gamma$ with the ATLAS Detector, Phys. Rev. Lett. **125**, 061802 (2020).

[15] CMS Collaboration, Measurement of the $t\bar{t}b\bar{b}$ production cross section in the all-jet final state in pp collisions at $\sqrt{s} = 13$ TeV, Phys. Lett. B **803**, 135285 (2020).

[16] J. Erdmann, S. Guindon, K. Kroeninger, B. Lemmer, O. Nackenhorst, A. Quadt, and P. Stolte, A likelihood-based reconstruction algorithm for top-quark pairs and the KLFitter framework, Nucl. Instrum. Methods Phys. Res., Sect. A **748**, 18 (2014).

[17] D. Pfau, J. Spencer, A. de G. Matthews, and W. Foulkes, Ab-initio solution of the many-electron Schrödinger equation with deep neural networks, Phys. Rev. Research **2**, 033429 (2020).

[18] J. S. H. Lee, I. Park, I. J. Watson, and S. Yang, Zero-permutation jet-parton assignment using a self-attention network, arXiv:2012.03542.

[19] A. Bogatskiy, B. Anderson, J. T. Offermann, M. Roussi, D. W. Miller, and R. Kondor, Lorentz group equivariant neural network for particle physics, arXiv:2006.04780.

[20] P. Baldi, The inner and outer approaches for the design of recursive neural networks architectures, Data Min. Knowl. Disc. **32**, 218 (2018).

[21] P. Baldi, *Deep Learning in Science: Theory, Algorithms, and Applications* (Cambridge University Press, Cambridge, England, 2020).

[22] F. Agostinelli, S. McAleer, A. Shmakov, and P. Baldi, Solving the Rubik's cube with deep reinforcement learning and search, Nat. Mach. Intell. **1**, 356 (2019).

[23] T. Cohen, M. Weiler, B. Kicanaoglu, and M. Welling, *Gauge Equivariant Convolutional Networks and the Icosahedral CNN* (PMLR, Long Beach, California, USA, 2019), pp. 1321–1330.

[24] T. Cohenand and M. Welling, *Group Equivariant Convolutional Networks* (PMLR, New York, New York, USA, 2016), pp. 2990–2999.

[25] J. Lee, Y. Lee, J. Kim, A. Kosiorek, S. Choi, and Y. W. Teh, *Set Transformer: A Framework for Attention-Based Permutation-Invariant Neural Networks* (PMLR, Long Beach, California, USA, 2019), pp. 3744–3753.

[26] M. Zaheer, S. Kottur, S. Ravanbakhsh, B. Poczos, R. R. Salakhutdinov, and A. J. Smola, Deep sets, in *Advances in Neural Information Processing Systems 30*, edited by I. Guyon, U. V. Luxburg, S. Bengio, H. Wallach, R. Fergus, S. Vishwanathan, and R. Garnett (Springer Science and Business Media LLC, Berlin, Germany, 2017), pp. 3391–3401.

[27] P. T. Komiske, E. M. Metodiev, and J. Thaler, Energy flow networks: deep sets for particle jets, J. High Energy Phys. 01 (2019) 121.

[28] Z. Wu, S. Pan, F. Chen, G. Long, C. Zhang, and P. S. Yu, A comprehensive survey on graph neural networks, IEEE Trans. Neural Netw. Learn. Syst. **32**, 4 (2021).

[29] J. Shlomi, P. Battaglia, and J.-R. Vlimant, Graph neural networks in particle physics, Mach. Learn. **2**, 021001 (2021).

[30] A. Vaswani, N. Shazeer, N. Parmar, J. Uszkoreit, L. Jones, A. N. Gomez, L. u. Kaiser, and I. Polosukhin, Attention is all you need, in *Advances in Neural Information Processing Systems 30*, edited by I. Guyon, U. V. Luxburg, S. Bengio, H. Wallach, R. Fergus, S. Vishwanathan, and R. Garnett (Curran Associates, Inc., Red Hook, New York, 2017), pp. 5998–6008.

[31] J. Devlin, M.-W. Chang, K. Lee, and K. Toutanova, BERT: Pre-training of deep bidirectional transformers for language understanding, in *Proceedings of the 2019 Conference of the North American Chapter of the Association for Computational Linguistics: Human Language Technologies, Volume 1 (Long and Short Papers)* (Association for Computational Linguistics, Minneapolis, Minnesota, 2019), pp. 4171–4186.

[32] A. Radford, J. Wu, R. Child, D. Luan, D. Amodei, and I. Sutskever, Language models are unsupervised multitask learners (2019).

[33] J. Alwall, R. Frederix, S. Frixione, V. Hirschi, F. Maltoni, O. Mattelaer, H. S. Shao, T. Stelzer, P. Torrielli, and M. Zaro, The automated computation of tree-level and next-to-leading order differential cross sections, and their matching to parton shower simulations, J. High Energy Phys. 07 (2014) 079.

[34] T. Sjöstrand, S. Ask, J. R. Christiansen, R. Corke, N. Desai, P. Ilten, S. Mrenna, S. Prestel, C. O. Rasmussen, and P. Z. Skands, An introduction to PYTHIA 8.2, Comput. Phys. Commun. **191**, 159 (2015).

[35] J. de Favereau, C. Delaere, P. Demin, A. Giammanco, V. Lemaitre, A. Mertens, and M. Selvaggi (DELPHES 3 Collaboration), DELPHES 3, A modular framework for fast simulation of a generic collider experiment, J. High Energy Phys. 02 (2014) 057.

[36] M. Cacciari, G. P. Salam, and G. Soyez, The anti-$k_t$ jet clustering algorithm, J. High Energy Phys. 04 (2008) 063.

[37] M. Cacciari, G. P. Salam, and G. Soyez, FASTJET user manual, Eur. Phys. J. C **72**, 1896 (2012).

[38] Spa-Net code release (2020), https://github.com/Alexanders101/SPA_Net_Code_Release.

[39] P. Gras, S. Höche, D. Kar, A. Larkoski, L. Lönnblad, S. Plätzer, A. Siódmok, P. Skands, G. Soyez, and J. Thaler, Systematics of quark/gluon tagging, J. High Energy Phys. 07 (2017) 091.

[40] ATLAS Collaboration, Performance of top-quark and $W$-boson tagging with ATLAS in Run 2 of the LHC, Eur. Phys. J. C **79**, 375 (2019).

[41] CMS Collaboration, Identification of heavy, energetic, hadronically decaying particles using machine-learning techniques, J. Instrum. **15**, P06005 (2020).

[42] I. Loshchilov and F. Hutter, Decoupled weight decay regularization, in *International Conference on Learning Representations* (2019).

[43] A. J. Larkoski, I. Moult, and B. Nachman, Jet substructure at the Large Hadron Collider: A review of recent advances in theory and machine learning, Phys. Rep. **841**, 1 (2020).

[44] ATLAS Collaboration, ATLAS b-jet identification performance and efficiency measurement with $t\bar{t}$ events in pp collisions at $\sqrt{s} = 13$ TeV, Eur. Phys. J. C **79**, 970 (2019).

[45] T. S. Caetano, J. J. McAuley, L. Cheng, Q. V. Le, and A. J. Smola, Learning graph matching, arXiv:0806.2890.

[46] B. Bloem-Reddyand and Y. W. Teh, Probabilistic symmetries and invariant neural networks, J. Mach. Learn. Res. **21**, 2 (2020).

[47] K. He, X. Zhang, S. Ren, and J. Sun, Delving deep into rectifiers: Surpassing human-level performance on imagenet classification, in *Proceedings of the 2015 IEEE International Conference on Computer Vision (ICCV)* (2015), pp. 1026–1034.

[48] S. Ioffeand and C. Szegedy, Batch normalization: Accelerating deep network training by reducing internal covariate shift, in *Proceedings of the 32nd International Conference on International Conference on Machine Learning— Volume 37*, ICML'15 (2015), pp. 448–456.

[49] J. L. Ba, J. R. Kiros, and G. E. Hinton, Layer normalization, arXiv:1607.06450.

[50] D. G. A. Smithand and J. Gray, opt_einsum—a PYTHON package for optimizing contraction order for einsum-like expressions, J. Open Source Software **3**, 753 (2018).

[51] L. Hertel, J. Collado, P. Sadowski, J. Ott, and P. Baldi, Sherpa: Robust hyperparameter optimization for machine learning, SoftwareX **12**, 100591 (2020).